\documentclass[letterpaper]{article}
  \usepackage{amsmath}
  \usepackage{graphics} 
  \usepackage{epsfig} 
  \usepackage{epstopdf}

\usepackage[latin1]{inputenc}
\usepackage{minitoc}
  \usepackage[colorlinks=true]{hyperref} 

\newcommand{\Proxm}{P_{\alpha}}

\title{The Reconstruction of Science Phylogeny\footnote{Both authors have equally contributed to this work.}
}

\begin{document}

\maketitle

\centerline{\scshape David Chavalarias }
\medskip
{\footnotesize
  \centerline{Institut des Systmes Complexes de Paris Ile-de-France \& CREA, CNRS - Ecole Polytechnique}
 \centerline{ ISCPIF, 57-59 rue Lhomond, 75005, Paris, France}
  \centerline{ http://chavalarias.com}
  } %

\medskip

\centerline{\scshape Jean-Philippe Cointet }
\medskip
{\footnotesize
 \centerline{CREA, CNRS - Ecole Polytechnique, 1 rue Descartes, 75005, Paris, France}
 \centerline{TSV, INRA, 65 av de Brandebourg, 94205, Ivry, France}
  \centerline{ http://jph.cointet.free.fr}
} 

\bigskip
\maketitle

\begin{abstract}
We are facing a real challenge when coping with the continuous acceleration of scientific production and the increasingly changing nature of science. In this article, we extend the classical framework of co-word analysis to the study of scientific landscape evolution. Capitalizing on formerly introduced science mapping methods with overlapping clustering, we propose methods to reconstruct \emph{phylogenetic networks} from 
successive science maps, and give insight into the various dynamics of scientific domains.
Two indexes -  the \emph{pseudo-inclusion} and the \emph{empirical quality} - are introduced to qualify scientific fields and are used for reconstruction validation purpose.
Phylogenetic dynamics appear to be strongly correlated to these two indexes, and to a weaker extent, to a third one previously introduced (\emph{density index}). These results suggest that there exist regular  patterns in the ``life cycle'' of scientific fields. The reconstruction of science phylogeny should improve our global understanding of science evolution and pave the way toward the development of innovative tools for our daily interactions with its productions. Over the long run, these methods should lead quantitative epistemology up to the point to corroborate or falsify theoretical models of science evolution based on large-scale phylogeny reconstruction from databases of scientific literature.
\end{abstract}


\bigskip


We are facing a real challenge when coping with the increasingly changing nature of science. First, the millions of papers published every year make clearly impossible for anybody to be aware of all the important breakthroughs and developments in science. This issue is made even more critical by the continuous acceleration of scientific production, which threatens every scholar with \textit{information overload } (the volume of  publications per year has doubled the last 12 years). Second, although science is not carved in marble and would better be defined as an ever-changing enterprise \cite{Hull-1988},  a lively debate has been taken place for more than 10 years around the shift toward a new regime of  knowledge production  following the transformation of the nature of the research process. According to \cite{nowotny2001rts} science has recently entered  a new mode, where knowledge is generated within a wider context of application, making full place to trans-disciplinarity, defined as the circulation of tools, theoretical perspectives, and people.
Whatever the causes of such transformations, the frontiers of science indeed appear to be  even faster changing and getting blurred as fields and sub-fields are cross-fertilizing, growing or dying. There is an urge to \textit{map these fluctuating landscapes}. 




Science mapping is one of the aims of scientometrics, a young science that took off in the late seventies, fostered by the development of electronic scientific databases and the increasing power of computers. Data-mining methods (in the wide sense) have been developed that make it possible to identify patterns, or \textit{meso structures} in scientific corpora that make sense to us (\textit{e.g.}   \emph{scientific fields} or \emph{epistemic fields}). The articulation between these scientific fields are then displayed on science maps to give overviews of scientific domains.

Part of the utility of science maps, both for theorists (science studies, history and philosophy of science), for users (scientists) or policy makers, comes from their capacity to give meaning to the evolution of science: what are the emergent fields, the continuities and main paradigmatic shifts, and from which scientific fields does a new field inherit its intellectual background. There is thus an important concern about reconstructing these dynamics in  such a way that fields of knowledge could be tracked through time. From the theoretical point of view, this entails that the core object in the representation of the evolution of science is a \emph{phylogenetic network} while most scientometrics studies focus on science snapshots. In this article,  we will show that co-word analysis is a suitable approach from this perspective and propose methods for an automated reconstruction of science phylogenies. The core question is: \textit{How can we reconstruct science dynamics through automated bottom-up analysis of scientific publications? }

%

 \section{Science mapping}
A large proportion of science maps are built upon co-occurrence data, with the assumption that the more likely two elements co-occur in the same article, the more they are related, and the closer they should appear on the map. These co-occurrence data can be of different nature: co-authorship networks, \cite{newman2004who}, co-citation networks, \cite{Small1973Citations1} or co-word networks (\cite{Callon1983From}, \cite{Callon1986Mapping}). In what follows, we will focus on these latter in the framework of co-word analysis. In this approach, co-occurrences of terms are indexed in large corpora. A graph structure is then generated, where nodes represent the terms, and strength of links represents their alleged similarity. This similarity measure is computed from co-occurrences data.  Higher level structures reflecting domains of science are then derived by analyzing patterns in this graph with clustering methods.

Scientometrics has defined a great number of measures based on co-occurrence data that capture the degree of similarity or proximity between two terms (\textit{cf.} \cite{He1999Knowledge} for a good review). Among others, we can mention two indexes that have been introduced early in scientometrics: the  inclusion index $\frac{n_{ij}}{min(n_i,n_j)}$ and the proximity index $\frac{n_{ij}^2}{n_i.n_j}$ \cite{Callon1986Qualitative}. Here, $n_i$ (respectively $n_j$ and $n_{ij})$ is the number of articles mentioning the term $i$ (respectively $j$ and both $i$ and $j$).

Further measures where later introduced. However, most of them, by synthesizing the relation between two terms with a single number, fail to convey important information about their use: given two terms $i$ and $j$, is one more specific or more generic than the other? Is $i$ more specific in the sense that it tends to be used by a sub-community of the community using  $j$?

We assume that the asymmetrical relation between terms is an essential information to get insight into the overall structure of science (fields and subfields). It can be captured by an appropriate choice of proximity measure such that  the \textit{pseudo-inclusion measure} defined over a period $T$ by\footnote{$n_i^T$ (resp. $n_j^T$ and $n_{ij}^T)$ is the number of articles mentioning the term $i$ (resp. $j$ and both $i$ and $j$) over the period $T$.}: $\Proxm^T(i,j)=((\frac{n_{ij}^T}{n_i^T})^{\alpha}(\frac{n_{ij}^T}{n_j^T})^{1/\alpha})^{min(\alpha,\frac{1}{\alpha})}$.

This measure has the advantage to convey information about the relative position of two terms from the point of view of their use: terms $j$ such that $\Proxm^T(i,j)$ is close to $1$ will contextualize $i$ for $\alpha \gg 1$  and  will tend to be more specific in their use relatively to $i$  for $0<\alpha \ll 1$ (see \cite{chava:scien} for more details)\footnote{Note that $\Proxm^T(i,j)=P_{\frac{1}{\alpha}}^T(j,i)$ so that if $j$ specifies $i$, $i$ contextualizes $j$. Moreover, $lim_{\alpha \rightarrow \infty}(\Proxm(i,j))$ is the inclusion measure over the sets of papers mentioning $i$ and $j$.}.

The pseudo-inclusion measure also enables a natural representation of the internal structure of a cluster $C$. To each term $w$ in $C$,  two coordinates $(I_s^\alpha(w),I_g^\alpha(w))$ can be assigned to qualify its degree of specificity and genericity relatively to other terms in $C$. \textit{The specificity index} indicates to what extent $w$ is specific to $C$ and is defined by: $I_s^\alpha(w)=\frac{1}{card(C)}\sum_{w'\in C} P_{max(\alpha,\frac{1}{\alpha})}(w,w')$. The \textit{genericity index} indicates to what extent a term $w$ contextualizes $C$. It is defined by: $I_g^\alpha(w)=\frac{1}{card(C)}\sum_{w'\in C}P_{min(\alpha,\frac{1}{\alpha})}(w,w')$. With this representation, the labeling  of each cluster finds a natural solution since each of its component is characterized on a specificity / genericity scale. According to what  is looking for, one can label the clusters with its most generic terms, its most specific ones, an so on (see \cite{coint08multi} for more details).


Starting from a set of terms $\mathcal{L}$ to be mapped (see the material and methods for the selection of terms and their indexation),  the pseudo-inclusion measure transforms the co-occurrence matrix  into an asymmetric proximity matrix $\mathcal{P}_\alpha$. This matrix defines a directed weighted graph on  $\mathcal{L}$ that can be further analyzed with clustering methods to detect informative patterns. In our case, patterns will represent domains of science defined by sets of strongly related terms that contextualize each other's meaning, some being more specific, others more generic. These sets will be called thereafter \emph{scientific fields}.

Several clustering methods have been proposed in literature and extensively tested for science mapping, \textit{e.g.} k-means clustering (\cite{Zitt2006Delineating}, \cite{boyack2005mapping}), Self-Organized Maps \cite{Skupin:2004p2187}, information flows based \cite{Rosvall2008Maps}. However,  terms can be used by different scientific communities with different meanings. This implies that some terms could belong to different scientific fields, a fact which technically requires the use of clustering methods allowing clusters overlap\footnote{
 \cite{Skupin:2004p2187} allows for the same label to belong to several knowledge domains, yet SOM  methods used to categorize abstracts indeed perform a partitioning.}. In order to keep the information conveyed by the asymmetry of $\mathcal{P}$ and allow  clusters overlap, we choose to consider the detection of directed cliques \cite{palla:dir} as basis for our clustering algorithm.  Extraction of directed cliques is one of the recent and convincing algorithm that produces overlapping clusters on directed graphs. 
In what follows, the set of directed cliques (or scientific fields) is noted $\mathcal{C} = \{C_i\}_{i \in I}$.

After this first clustering operation, the next step is to give an insight into the articulation of the different scientific fields to provide a global view of the scientific landscape covered by $\mathcal{L}$.

The pseudo-inclusion measure $\Proxm$ can naturally be extended to proximity between clusters at period $T$ by averaging the proximity between terms of two clusters:

$$ \Proxm^{T,2}(C_a,C_b)=\frac{1}{\mid C_a \mid} \sum_{i \in C_a}(\frac{1}{\mid C_b \mid}\sum_{j\in C_b}\Proxm^T(i,j))\label{intercluster}$$

It is important to note that two clusters can be close relatively to $ \Proxm^{T,2}$ even if they do not share any terms from the moment the terms they contain are related.

$\Proxm^{T,2}$ defines a weighted directed graph on the set of clusters that can be mapped with network visualization tools. Automatic cluster labeling can profitably be exploited to further simply the map by merging clusters with same labels. Depending on the labeling chosen (specific labels, generic, etc.) and the number of labels per cluster, visualizations will display different view points on the scientific domain under study, with different resolutions.

\section{Validation}
As stated before, the aim of phylogeny reconstruction is to discover patterns and regularities in science evolution. Given this objective, we defined two benchmarks for this reconstruction: theoretical validation and empirical validation.

Theoretical validation is related to  the robustness of the detected patterns regarding the dataset (\cite{Hopcroft2004Tracking}) and the parameters of the model ($d_0$ in our case). Detected patterns should be robust to parameter change if we want them to be significant.

Empirical validation is related to the adequacy of the reconstruction of scientific fields compared to the actual productions of scientific communities. To reflect the activity of a scientific community, it is important that scientific fields be composed with terms that are indeed mentioned altogether in the literature. The principle of the proposed empirical validation is thus to check, for each cluster, that there is some significant number of papers mentioning all the terms of the clusters in their full text. Moreover, a cluster composed by very common terms (\textit{e.g.} {disease ,molecule,cell,division}) are not as much informative as a cluster composed of more specific terms (\textit{e.g.} {cancer ,dna damage, apoptosis, checkpoint}). This nuance can be caught by the notion of self-information \cite{shannon1948mathematical} conveyed by the observation of an event composed of independent items $a_1$ ... $a_n$ which have a probability $p_1$ ... $p_n$ to be observed individually. Self-Information is then defined by $I(a_1,...,a_n)=\sum_{i=1...n}-log(p_i)$. These two constraints can be synthesized into the \textit{empirical quality} of a cluster $C$, defined as the products of its self-information with the normalized number $\frac{n_C}{N}$ of papers mentioning all the terms of $C$ in their full text: $Q_e(C)= \frac{n_C}{N}.\sum_{i \in C}-log(\frac{n_i}{N})$, where $N$  is the total number of papers in the reference corpus.
The empirical quality could be used as a parameter to filter phylogenies so as to display most relevant scientific fields.

\section{Qualifying clusters}
Relevance is not a binary judgment but rather lays on a continuum, potentially multidimensional, reflecting what is looked for:  well-recognized domains of investigation, emergent domains, highlights on interdisciplinary domains, etc. Empirical quality is one of the indexes that make it possible to qualify identified scientific fields. Furthermore, we studied  two other indexes  that help to give meaning to science evolution.
\begin{itemize}
\item \textbf{Density.} One of the first index introduced to assess scientific fields evolution is the density of a field \cite{callon91coword}.``It characterizes the strength of the links that tie the words making up the cluster together. The stronger these links are, the more the research problems corresponding to the cluster constitute a coherent and integrated whole. It could be said that density provides a good representation of the cluster's capacity to maintain itself and to develop over the course of time in the field under consideration." It is computed by: $D(C)=\frac{1}{Card(C)}\sum_{(w,w')\in C^2, w\neq w'} P_1(w,w')$,
\item \textbf{Pseudo-inclusion index.} 
    Since our goal is to find clusters where all terms are satisfying contexts or well contextualized by other terms in the cluster, we defined the \emph{pseudo-inclusion index} of a cluster: $I_{\subset}^\alpha(C)=\min_{w \in C}\frac{1}{2}(I_s^\alpha(w)+I_g^\alpha(w))$. This index indicates the degree of structuration of $C$. Clusters with low pseudo-inclusion index have at least one term that does not fit well with other terms, being neither specific nor generic. As we shall see, the pseudo-inclusion opens some perspectives to the interpretation of science dynamics.
\end{itemize}

Along with empirical quality, these two indexes will be useful to filter science maps and focus on some particular parts of the phylogeny. Note that whereas pseudo-inclusion and density  can be computed without supplementary information, empirical quality needs additional queries to a corpus database. One issue will thus be to see the extent to which it is possible to use the first two indexes as proxies to evaluate the empirical quality.

\section{Tracking meso-dynamics}
One of the most essential features of the evolution of science  is the way in which new associations between terms are performed and change the composition of scientific fields. These changes in the use of terms are the main visible evidences of shifts in scientific activity. Sets of terms are the adequate level to study cross-fertilization of different fields of science, circulation of concepts through domains, bursts of activity in a given branch, and so on. They are widely  used by scientists, to define with few keywords, their research,  a journal topics or a conference scope.  We will call the dynamics of science studied at the level of sets of terms the \textit{meso-dynamics} of science. Reconstructing these meso-dynamics is equivalent to finding a matching function between clusters of science maps between successive periods of time.

 The answer to this problem is far from straightforward.  A scientific field, represented by a cluster $C$ at a given period of time, can undertake several kinds of transformation in its composition that will entails a different representation in the next periods: it can gain new concepts, loose others, merge with an other field, split or die. Consequently, two successive maps can have very different sets of scientific fields. However, even if scientific fields were all different between two periods, they could nevertheless share some terms and potentially share a common scientific background. A scientific field can have several ``offsprings'' in the next period and its conceptual legacy may come from several domains of investigation from the previous period.

 The reconstruction of these inheritance patterns will be very useful to get a global overview of the activity and evolution of large scientific domains. Moreover, contrary to what is often encountered in biology, we should expect some hybridization events between fields of research, which requires switching from phylogenetic trees to \textit{phylogenetic networks}. Reconstructing the phylogenetic network of science consists in answering this simple question: given a scientific field $C^{T'}$ at period $T'$ and a period $T$ prior to $T'$, from which fields at $T$ does $C^{T'}$ derives its conceptual legacy?

To achieve inter-temporal matching between fields, we have to find for each field at $T$ the field or union of fields from which it inherits.  We assume that the time scale of the evolution scientific fields  is slow enough to allow simple similarity measures between two close periods to track  the meso-dynamics of a given field. We thus  seek to find the field \emph{or} combination of fields that are most similar and therefore the most likely matchable. One of the most straightforward measure is a Jaccard similarity measure\footnote{This function is the inverse of the ``transformation index'' introduced for similar purposes by Callon in \cite{callon91coword}} on fields terms, thereafter denoted $d$. Given two fields $C_1$ and $C_2$, $d(C_1,C_2)=\frac{| C_1 \cap  C_2 |}{|  C_1 \cup  C_2 |} $.  $d$ can be interpreted in terms of  the probability that a term belonging to $C_1 \cup  C_2 $ also belong to $C_1 \cap  C_2$. This is simply a measure of the overlap between $C_1$ and $C_2$.

Given a conceptual field $C_l^{T'}\in\{C_b^{T'}\}_{b \in B}$ at time $T'$, we propose to perform inter-temporal  matching by choosing its ``fathers'' $\Phi_l^{T'}$ among the set of paradigmatic fields of the previous period $\{C_a^T\}_{a \in A}$ as: $$ \Phi^{T'}(C_l) = argmax_{K \subset A} (d({\bigcup_{k \in K} C^{T}_k ,C^{T'}_l }))$$

With the Jaccard similarity measure we can write:
$$ \Phi^{T'}(C_l) = argmax_{K \subset  A}\frac{|(\cup_{k \in K} \mathcal{C}^{T}_k ) \bigcap C^{T'}_l  |}{| (\cup_{k \in K} C^{T}_k ) \bigcup  C^{T'}_l |}$$

  \begin{figure}
\centerline{\includegraphics[width= 5cm]{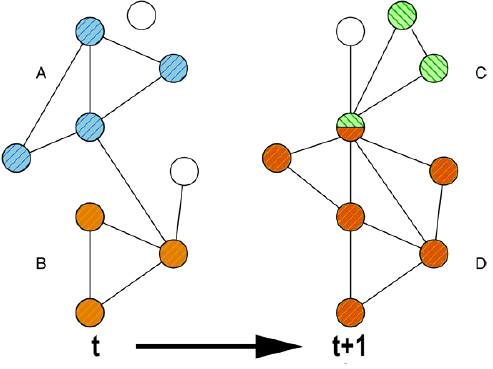}}
\caption{Inter-temporal fields matching.}\label{figure1}
\end{figure}

 Figure 1.  illustrates the matching procedure. We plotted two successive sub-networks with the same set of nodes  between two  time steps. The two successive period present distinct cluster sets : $A$ and $B$ at time $t$ and $C$ and $D$ at time $t+1$. Note that one node  belongs to two different clusters at time $t+1$. The aim is to determine from which fields or union of fields $C$ and $D$ may be descending. It is straightforward to check that field $A$ is the closest to cluster $C$ (\emph{i.e.} $\Phi^{t+1}(C) = A$). Even if two nodes were removed from $A$
 while one node was added, the similarity between $A$ and $C$ ($d(A,C)=\frac{2}{5}$)  is still the  best possible and offers the best matching. The case of $D$ is more delicate since three cases are possible: $D$ may inherit from $A$, $B$ or $A\cup B$. Computing the distances according to each cases we get: $d(D,A) = \frac{2}{8}$, $d(D,B)=\frac{3}{6}$ and finally $d(D,A \cup B)=\frac{5}{7}$. We will thus conclude that $D$ most likely inherits from  the merging of the two preceding fields $A$ and $B$ and thus conclude that $\Phi^{t+1}(D) = A\cup B$.

Since it would seem incorrect to match two fields that have very few terms in common even though no better matching is possible, we need to define a threshold above which the matching is satisfying. We shall call this threshold $d_0$. One can tune this threshold requiring a minimum amount of similarity. As we shall see, activity patterns in the phylogeny (areas of activity burst, areas with emergent fields, branches death, etc.) are robust to variations of $d_0$ provided that $d_0$ does not get too close from 0 or 1.

\section{Phylogenetic Patterns}
We performed phylogeny reconstruction on the MedLine database focusing on research in biological and biomedical fields related with network studies. After the constitution of a database concerning a set $\mathcal{L}$ of 834 terms (\textit{cf. } materials and methods), we generated maps processed on four years sliding time windows from 2007 to 1987. An exemple of these maps is given in figure 2.

\begin{figure}
\centering
  \includegraphics[width=13cm]{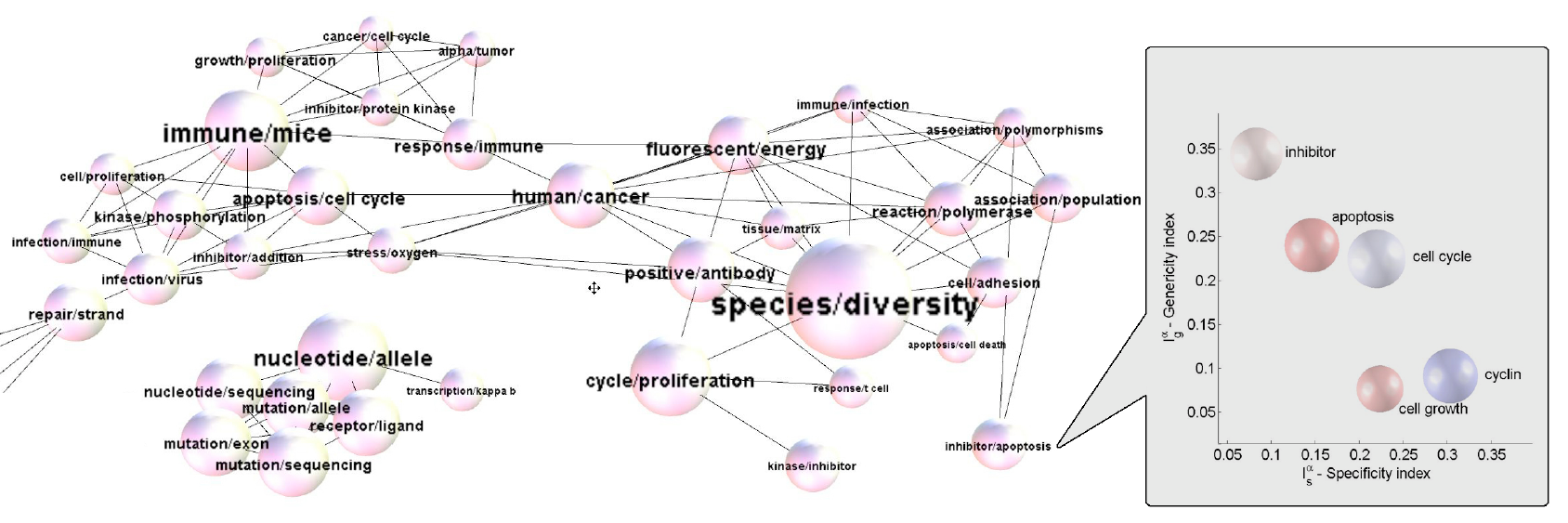}\
  \caption{Detail of the map related to network studies in biology obtained on the period 2004-2007 for terms list $\mathcal{L}$ (see material and methods). Clusters were required to have at least 4 terms. Clusters are labeled with their two most generic terms and merged with clusters with the same label. Size of text and bubbles map the density of clusters. This value is averaged over the set of merged clusters if the node is made of merged clusters. The value of the link between two sets of merged clusters is the maximum value of the inter-cluster similarity between all pairs of clusters \tiny (Visualized with Gephi.org). \normalsize The inset gives a detail of the cluster labeled \textit{inhibitor/apoptosis} plotted in the $(I_s^\alpha(w),I_g^\alpha(w))$ space. Sizes of the bubbles map the number of co-occurrences with other terms of the cluster and colors these numbers growth rates compared to 2004-2007. }\label{sciencemap}
  \label{figure3}
\end{figure}

We reconstructed the phylogeny of the domains related to networks studies in biology over the period 1987-2007 and studied the patterns of three indexes of cluster structuration: the density, pseudo-inclusion and empirical quality. Releasing all constraints on the phylogeny except that we required the fields to have at least four elements and a non null empirical quality, the phylogenetic  network obtained was made of 7759 nodes.

Within  this network, we observed a significant positive correlation between the pseudo-inclusion index and the empirical quality. The Pearson coefficient $r$ lays within the 95\% confidence interval $[0.14;0.19]$, the probability to obtain a correlation as large as the observed value by random chance being $p=4.10^{-39}$. Between the pseudo-inclusion index and the number of papers per cluster we get $0.28<r<0.32$ and $p=0$. To a lesser extent, there is a significant positive correlation between the density index and the empirical quality ($0.03<r<0.08$, $p=4.10^{-6}$) as well as with the number of papers per cluster ($0.16<r<0.21$, $p=0$).

\begin{figure}
\centerline{\includegraphics[width=8cm]{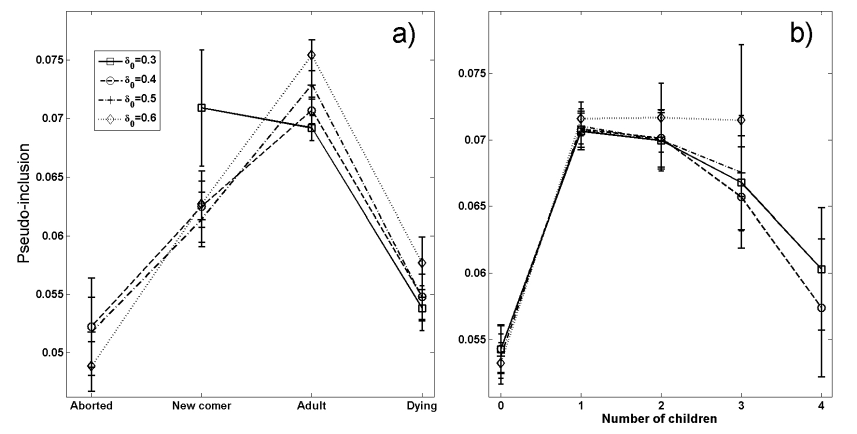}}
\caption{Dependencies of the mean of the pseudo-inclusionover the position of the fields in the phylogeny  (a) as well as over its number of sons (b) suggest trends in the ``life cycle'' of scientific fields: these indexes grow while a new field emerges in bushy branches, and then loose their strength when it begins to be neglected by the community. As shown, these patterns are robust against variations in the domain $0.3\leq d_0 \leq0.6$. Error bars indicate the 95\% confidence interval.}\label{figure2}
\end{figure}

We categorized the fields according their position in the phylogenetic network: aborted (no father, no child), new comers (no father, some children), adult (with father(s) and son(s))  and dying fields (with some father(s) but no child). Note that  a cluster may belong to a different category according to the value of $d_0$. The distribution of scientific fields regarding to these categories is particularly interesting. Plotting the variations of the fields' pseudo-inclusion (fig. 3.a), density and pseudo-inclusion (Appendix.1) indexes against this categorization, we found very clear patterns in the domain $0.3\leq d_0 \leq0.6$: aborted, new comer and dying fields tend to have weaker indexes than adult fields, with aborted fields having slightly lower values for their indexes than new comers. Similar patterns have been obtained for the density and empirical quality (\textit{cf.} Appendix.1).

The dependency of the mean of the density, pseudo-inclusion and empirical quality indexes over the position of the fields in the phylogeny suggests trends in the ``life cycle'' of scientific fields: these indexes grow while a new field emerges, and then loose their strength when it begins to be neglected by the community. However, density and pseudo-inclusion index are completely different ways of characterizing scientific fields.  On the one hand,  fields with high pseudo-inclusion will usually have terms with a large spectrum of specificity and genericity,  which means that they are likely to contain very specific terms with few occurrences. These terms have a high probability to be new concepts or new objects of study. Their presence in the phylogeny will then be correlated with high rate of branching processes. On the other hand, fields with a high density index correspond to well structured scientific domains  with a priori lower rate of conceptual renewal.

 Further studies based on different databases will confirm or not the relevance of these general patterns in the study of science evolution. However, these regularities open  perspectives for the detection of emergent or dying fields on the basis of some indexes computed on co-occurrence data.

Beside, the fact that aborted fields tend to be of lower quality suggests a methodology to adjust optimally $d_0$ in order to have the most informative phylogeny (in the sense of the empirical quality). Indeed, the ratio between the mean quality of fields belonging to the phylogeny and the mean quality of aborted fields is always higher than 1, and reaches its maximum around the value $d_c=0.33$. For this value, connected fields in the phylogeny \textit{i.e.} fields that have at least one father or one son, are on average almost twice as informative as aborted fields.


Inheritance patterns can be studied by classifying fields according to their number of sons in the phylogenetic network. While most fields have less than 2 sons, with 44\% having only one successor, almost 14\% have at least 3 children. Again, the  distribution of the different indexes in function of the number of children is very instructive. Figure~3.b shows that, on average, the maximum of density is obtained for fields that have only one son. Similar patterns have been obtained for the  pseudo-inclusion and the number of papers per cluster (\textit{cf.} Appendix.2). Again, this observation holds for a large range of $d_0$.
 The synthesis of all these results suggests that relatively young branches of science are generally bushy with fields having lots of children. This corresponds to an intense exploration of new directions of research. Older fields will generally have a much more linear evolution with a lower rate of conceptual renewal.
This pattern can clearly be observed on figure~4 that represents, for $d_0=d_c$, the subpart of the phylogenetic network composed of fields with highest empirical quality and at least four terms. Most recent branches have also been removed to meet editorial constraints. We can  also notice that there is much more hybridation between scientific fields in the  domain of formal methods and tools than in the branches corresponding to topics in biology. This transversal domain is also over-represented due to the fact that the targeted thematic (networks) is itself a transversal methodology.

  \begin{figure}
\centerline{\includegraphics[width=8cm]{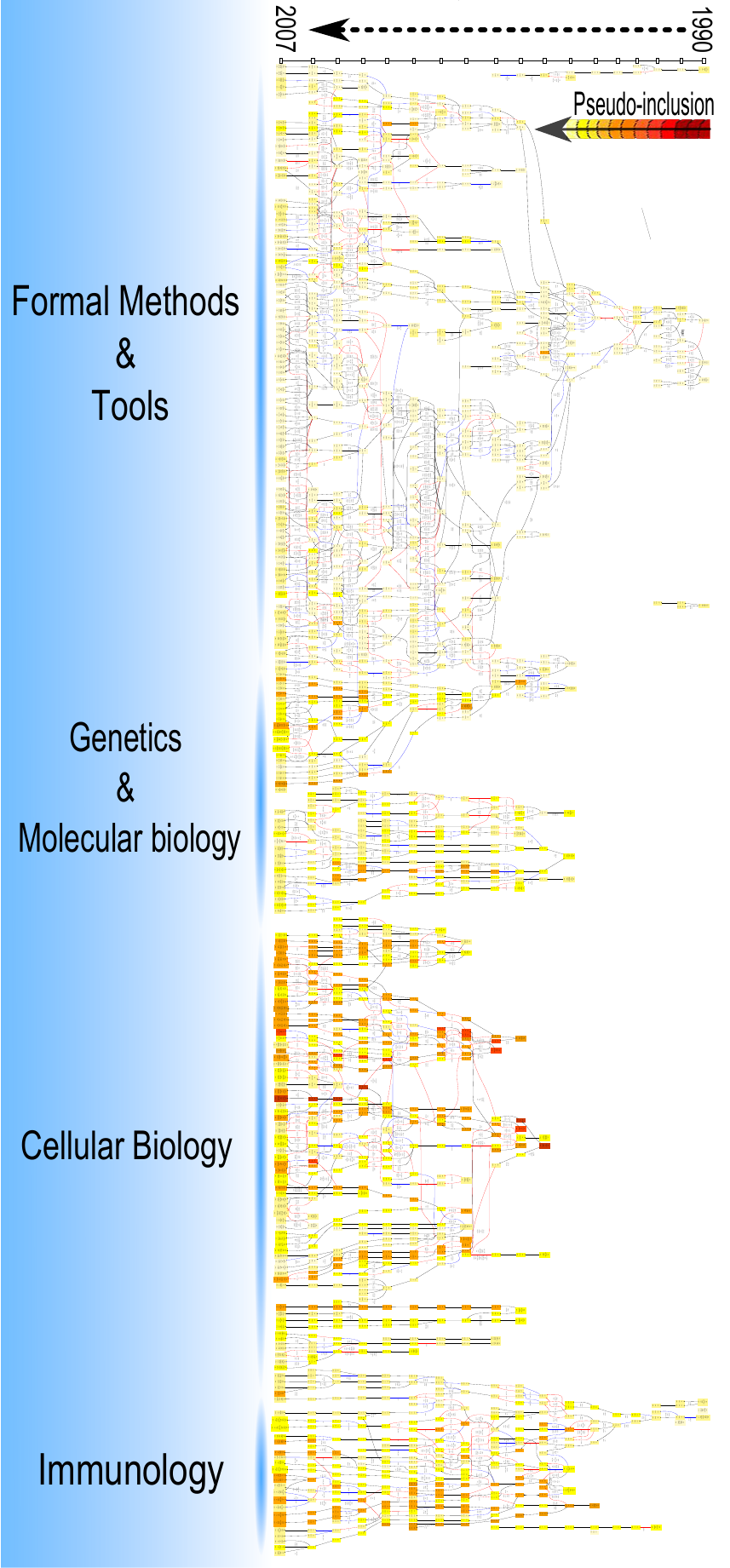}}
\caption{Extract  of the full phylogeny of domains related to networks studies in biology and medical research. We kept fields made of more than four terms, set a threshold on the empirical quality ($0.04$) and removed shortest branches for editorial purposes. Some branches have been gathered compared to GraphViz display on the basis of their thematic. Colors map the pseudo-inclusion index of the fields. Fields are labeled with their most generic term, except for the beginning of a branch or for the most recent period, where all terms are displayed. The labels of inter-period arrows indicate which terms have been lost or gained between two periods.The number on the first line of a field label is the field id, le number on the last line is the number of articles mentioning all terms of the fields in the reference database. Zoom in to see details.}\label{figure4}
\end{figure}

Details of the phylogeny are also very informative. Figure~5 represents the phylogeny with fields of more than five terms for which at least one term contains the words ``cancer'' or ``tumor''. On this partial phylogeny, we can clearly see three distinct sets of branches with very different characteristics. Two sets are quite bushy and deals with \emph{cancer }and \emph{DNA} issues on one side, \emph{cancer, tumor and proliferation} issues on the other side. They appear to have increased their interactions these last several years around the concepts of \emph{apoptosis}, \emph{suppressor} and \emph{cell cycle}. The third set has very linear branches and is related to the relations between \emph{tumor} and the \emph{immune system}. These three sets are also quite distinct in terms of the range of their density and pseudo-inclusion indexes. Whereas the bushy branches tend to have a higher pseudo-inclusion index than the linear ones, revealing  a higher rate of conceptual renewal, they also have a lower density index, indicating that they should be more recent. The study of the evolution of the pseudo-inclusion index along these branches reveals that this index is increasing along most of the branches although its growth rate is decreasing with time. When relaxing the constraints on the empirical quality threshold and on the number of terms in clusters, these characteristics regarding the three sets of branches are preserved, although  the branches prove to be older than they appear in this partial phylogeny, the upper-part of the phylogeny having been pruned in the thresholding process.

  \begin{figure}
\centerline{\includegraphics[width=12cm]{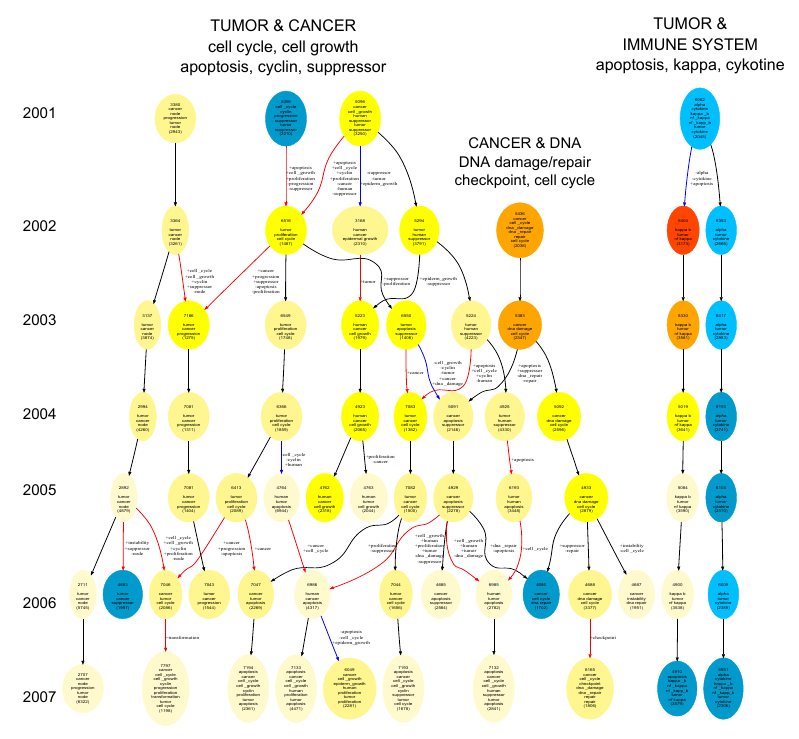}}
\caption{Detail of the sub-phylogenetic network related with cancer studies. Colors of the circles, from blue to red, maps the growth rate of the pseudo-inclusion index. Red links indicate the introduction of at least one new term. Note that this index is increasing along most of the branches (warm colors) although its growth rate is decreasing with time. Fields are labeled with their most generic term, except for the beginning of a branch or for the most recent period, where all terms are displayed. The labels of inter-period arrows indicate which terms have been lost or gained between two periods. In cluster labels, the number on first row indicates the cluster id and the number on last row  indicates the number of articles mentioning all terms of the cluster.}\label{figure3}
\end{figure}

\section{Toward quantitative epistemology}
The seminal work of Callon et. al. \cite{callon91coword} was the first attempt  to quantify the evolution of scientific fields through co-word analysis, monitoring inter alia, the evolution of the density of clusters. Our work proposes the first automated methods for the bottom-up reconstruction of the entire phylogeny of a domain of science and is clearly in line with their approach. We expanded their approach in several ways, trying to take into account the classical limitations of scientometrics that have been expressed hitherto.\\
\textbf{Coverage: } Co-word analysis can cover the largest bibliographic database available. Nowadays, online publishers  cover between 30 and 40 million articles, which represent a significant part of worldwide scientific literature. We gave an example on a case study based on MedLine (14M papers)  covering most medical and biological research.\\
\textbf{Ambiguity: } Contrary to \cite{callon91coword} and most subsequent works, we used overlapping clustering algorithms in order to ensure that we can handle ambiguity in terms use and avoid false negatives in scientific fields detection \emph{e.g.} terms that are classified in different clusters although they are strongly related.\\
\textbf{Asymmetry and bottom-up multi-level mapping: } Following previous work \cite{chava:scien}, we based our clustering algorithm on an asymmetric proximity measure  in order to fully reflect the organization of science into domains and sub-domains. This asymmetry makes it possible to highlight the internal structure of  clusters  allowing automatic labeling \cite{coint08multi} in a bottom-up way (similarly to \cite{Skupin:2004p2187} but contrary to top-down labeling, \textit{e.g.} \cite{Moya2004New}, \cite{boyack2005mapping} or \cite{Leydesdorff2008Dynamic} who use ISI journal classification to label clusters). This offers possibilities of multi-level mapping with multiple view points on the phylogeny according to the required degree of  specificity. We also introduced a measure of fields structuration, the \textit{pseudo-inclusion  index}, based on this new asymmetric proximity and we showed that the pseudo-inclusion index appears to be very informative when assessing the evolution of a fields of research.\\
\textbf{Validation: } Complementary to \cite{Heal86AnExp} who suggested to use both ``internal validation'' (\textit{i.e.} by experts of the domains) and ``external validation'' (\textit{i.e.} by users of the maps), and \cite{Hopcroft2004Tracking} who proposed a method to asses the stability of a clustering, we proposed an \textit{empirical validation} of  science maps (confrontation with real data) that complement these approaches. We introduced the \textit{empirical quality} that reflects the amount of information conveyed by a cluster about actual scientific activity and showed that the pseudo-inclusion index was positively correlated with the empirical quality. The density, on the other hand, was only weakly correlated.\\
\textbf{Dynamics: } The proposed methodology capitalises on the availability of diachronic data to reconstruct the phylogeny of scientific fields, and takes into account multiple filiations, contrary to what could have been done in other related fields like social group evolution \cite{Palla:2007p229} or \cite{Hopcroft2004Tracking}. The reconstructed science phylogeny revealed strong and robust patterns which appear to highlight strong regularities in science evolution.\\

This approach opens perspectives both from theoretical and applicative points of view. While we tried to show that researches in the reconstruction of science dynamics  are close to the point where they will  make it possible to corrobate or falsify theories in epistemology and science studies, we can also expect they will  considerably renew the way we interact with science, especially when browsing large-scale electronic databases. Moreover, the methodology presented here is not specific to scientific corpora and may be applied to a wide range of co-occurrence data from online communities, patents database, folksonomies, web queries or even experimental data like micro-array data.

\section{Appendix}
\subsection{Indexation: from corpus to data}
In order to propose scalable methods on rough data, we considered indexes of science databases as proxies to evolution of science, \textit{e.g.} as they are already built by search engines. Our method thus cope with the constraint of working with aggregated co-occurrence data of terms in articles. Other methods bring interesting complementary perspectives in epistemic communities dynamics but require a more detailed access to data sets (like author-based data for example \cite{Roth2006Lattice}). 

 Co-word analysis critically depends on the initial set of terms chosen for  the study and can be biased by the ``indexer effect'' (\cite{Whittaker1989Creativity}, \cite{Callon1986Putting}, \cite{He1999Knowledge}). This effect can have several origins: terms selected by the indexers are too general, specific terms have been omitted from the satisfactory list or the indexer puts the wrong emphasis, or even a mistaken emphasis in keywording. For the case study presented in this paper, we choose a semi-automatic method that takes advantage both of powerful automated parsing of large corpora and experts skills to minimize this effect. We also choose to index terms within abstracts or full text of articles rather than in keywords lists provided by publishers or authors.

The case study  presented in this article targets the question of \textit{networks} in medical and biological research. We choose PubMed-MedLine as data source since it covers most of the publications in biology (more than 17M references), while titles and abstracts of articles are freely available. We then choose few concepts related to network-based approaches (network, evolvable, evolvability, hub, feedback) and retrieved all the papers mentioning at least one of these terms in MedLine (about 2,4M references). We then indexed these 2,4M abstracts with date of publication and retrieved all  n-grams\footnote{Key phrases with exactly n terms.} with a number of occurrences higher than $100^\frac{1}{n}$ and $n\leq3$ over the whole period (\emph{e.g.}  the term \textit{protein interaction network} has to appear at least in 5 references to be included in our set of candidate keywords). Stop words were discarded. This list of terms was then checked by science historians to further discard uninformative terms, which finally lead to a set $\mathcal{L}$ of 834 terms (available at http://www.maps.sciencemapping.com/eprint/phylo/appendix3.txt).

These terms were then indexed from 1950 to 2008 in the 2,69M retrieved abstracts to build the co-occurrence array $\mathcal{M}_t$ of all co-occurrences for terms in $\mathcal{L}$ from 1950 to 2008. $\mathcal{M}_t(i,j)$ gives the number of articles published during the year $t$ which mentioned both terms $i$ and $j$ in their abstract.

\subsection{Software}
We developed and used the Words Evolution software (http://sciencemapping.com/WE) to process and visualize the phylogenies. This software is interfaced with network visualization tools like Gephi  or Graphviz as well as clustering softwares like Cfinder.

\subsection{Supplementary figures}

\begin{figure}[!h]
\centering
  \includegraphics[width=8cm]{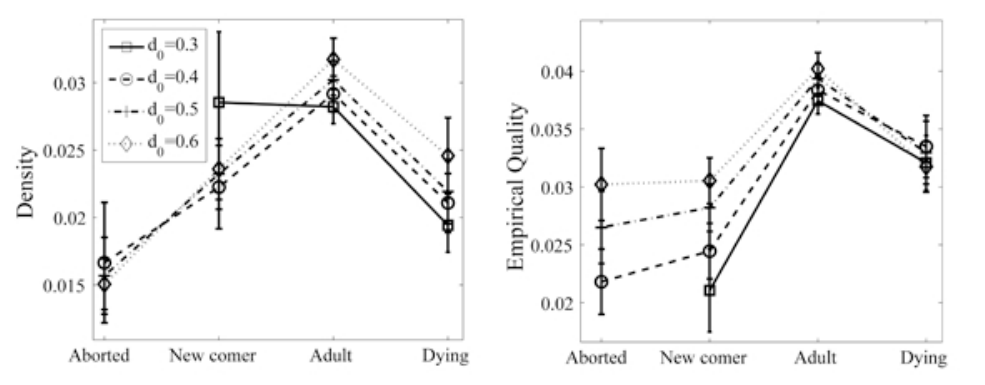}\
  \caption{Dependencies of the mean of the density and empirical quality over the position of the fields in the phylogeny. As shown, these patterns are robust against variations in the domain $0.3\leq d_0 \leq0.6$. Error bars indicate the 95\% confidence interval.}
  \label{SI1}
\end{figure}

\begin{figure}[!h]
\centering
  \includegraphics[width=8cm]{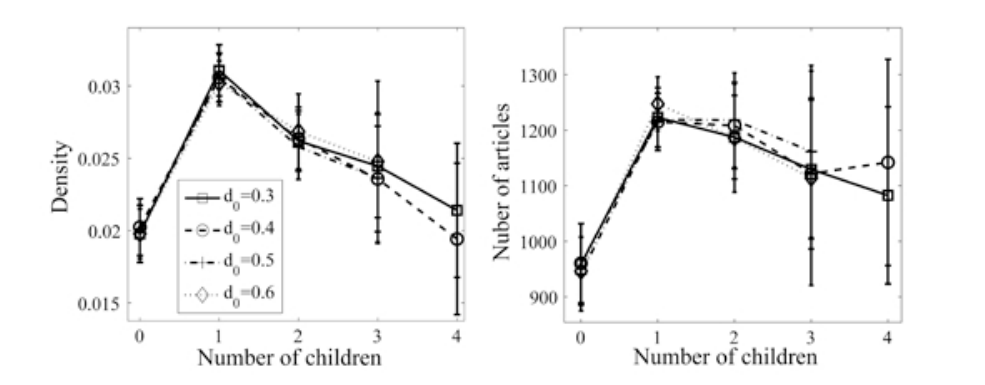}\
  \caption{Dependencies of the mean of the density of clusters and number of articles in fonction of the number of sons. As shown, these patterns are robust against variations in the domain $0.3\leq d_0 \leq0.6$. Error bars indicate the 95\% confidence interval.}
  \label{SI1}
\end{figure}

\section{acknowledgments}
The authors warmly thank Jean-Paul Gaudilli\`{e}re and Christophe Bonneuil for their help in selecting the list of terms. These researches have been supported by the FP7 PATRES project, the Paris \^{I}le-de-France Complex Systems Institute, the Ecole Polytechnique/CNRS, and the INRA.

\bibliography{phylogeny}

\begin{thebibliography}{10}

\bibitem{boyack2005mapping}
Kevin~W. Boyack, Richard Klavans, and Katy B\"{o}rner.
\newblock Mapping the backbone of science.
\newblock {\em Scientometrics}, 64(3):351--374, 2005.

\bibitem{callon91coword}
M.~Callon, J.~P. Courtial, and F.~Laville.
\newblock Co-word analysis as a tool for describing the network of interaction
  between basic and technological research: The case of polymer chemistry.
\newblock {\em Scientometric}, 22(1):155--205, 1991.

\bibitem{Callon1983From}
M.~Callon, J.~P. Courtial, W.~A. Turner, and S.~Bauin.
\newblock From translations to problematic networks: An introduction to co-word
  analysis.
\newblock {\em Social Science Information}, 22(2):191--235, 1983.

\bibitem{Callon1986Mapping}
M.~Callon, J.~Law, and A.~Rip.
\newblock {\em Mapping the Dynamics of Science and Technology}.
\newblock 1986.

\bibitem{Callon1986Putting}
M.~Callon, J.~Law, and A.~Rip.
\newblock {\em Putting texts in their place.}, pages 221--230.
\newblock London: The Macmillan Press Ltd., 1986.

\bibitem{Callon1986Qualitative}
M.~Callon, J.~Law, and A.~Rip.
\newblock {\em Qualitative scientometrics.}, pages 103--123.
\newblock London: The Macmillan Press Ltd., 1986.

\bibitem{chava:scien}
D.~Chavalarias and J.~P. Cointet.
\newblock Bottom-up scientific field detection for dynamical and hierarchical
  science mapping - methodology and case study.
\newblock {\em Scientometric}, 75(1):37--50, 2008.

\bibitem{coint08multi}
J.~P. Cointet and D.~Chavalarias.
\newblock Multi-level science mapping with asymmetric co-occurrence analysis:
  Methodology and case study,.
\newblock {\em Networks and Heterogeneous Media}, 3(2):267--276, June 2008.

\bibitem{He1999Knowledge}
Qin He.
\newblock Knowledge discovery through co-word analysis.
\newblock {\em Library Trends}, 48(1):133--159, 1999.

\bibitem{Heal86AnExp}
P.~Healey.
\newblock An experiment in science mapping for research planning.
\newblock {\em Research Policy}, 15(5):233--251, October 1986.

\bibitem{Hopcroft2004Tracking}
John Hopcroft, Omar Khan, Brian Kulis, and Bart Selman.
\newblock Tracking evolving communities in large linked networks.
\newblock {\em PNAS}, 101(1):5249--5253, April 2004.

\bibitem{Hull-1988}
D.~Hull.
\newblock {\em Science as a process: an evolutionary account of the social and
  conceptual development of science}.
\newblock Chicago: University of Chicago Press, 1988.

\bibitem{Leydesdorff2008Dynamic}
Loet Leydesdorff and Thomas Schank.
\newblock Dynamic animations of journal maps: Indicators of structural change
  and interdisciplinary developments.
\newblock {\em Journal of the American Society for Information Science and
  Technology}, 59(11):1810--1818, 2008.

\bibitem{Moya2004New}
F\'{e}lix Moya-Aneg\'{o}n, Benjam\'{i}n Vargas-Quesada, Victor Herrero-Solana,
  Zaida Chinchilla-Rodr\'{i}guez, Elena Corera-\'{A}lvarez, and Francisco
  Munoz-Fern\'{a}ndez.
\newblock A new technique for building maps of large scientific domains based
  on the cocitation of classes and categories.
\newblock {\em Scientometrics}, 61(1):129--145, September 2004.

\bibitem{newman2004who}
Mark Newman.
\newblock Who is the best connected scientist? a study of scientific
  coauthorship networks.
\newblock {\em Complex Networks}, pages 337--370, 2004.

\bibitem{nowotny2001rts}
H.~Nowotny, P.~Scott, and M.~Gibbons.
\newblock {Re-Thinking Science: Knowledge and the Public in an Age of
  Uncertainty}.
\newblock {\em Cambridge, Mass}, 2001.

\bibitem{Palla:2007p229}
Gergely Palla, Albert~L. Barab{\'{a}}si, and T.~Vicsek.
\newblock Quantifying social group evolution.
\newblock {\em Nature}, Jan 2007.

\bibitem{palla:dir}
Gergely Palla, Ill\'{e}s~J Farkas, P\'{e}ter Pollner, Imre Der\'{e}nyi, and
  Tam\'{a}s Vicsek.
\newblock Directed network modules.
\newblock {\em New Journal of Physics}, 9(6):186, 2007.

\bibitem{Rosvall2008Maps}
Martin Rosvall and Carl~T. Bergstrom.
\newblock Maps of random walks on complex networks reveal community structure.
\newblock {\em PNAS}, 105(4):1118--1123, January 2008.

\bibitem{Roth2006Lattice}
Roth, Camille, Bourgine, and Paul.
\newblock Lattice-based dynamic and overlapping taxonomies: The case of
  epistemic communities.
\newblock {\em Scientometrics}, 69(2):429--447, November 2006.

\bibitem{shannon1948mathematical}
Claude Shannon and Warren Weaver.
\newblock {\em Mathematical Theory of Communication}.
\newblock {University of Illinois Press}, June 2002.

\bibitem{Skupin:2004p2187}
A~Skupin.
\newblock The world of geography: Visualizing a knowledge domain with
  cartographic means.
\newblock {\em Proceedings of the National Academy of Sciences}, 101(Suppl
  1):5274--5278, Jan 2004.

\bibitem{Small1973Citations1}
H.~Small.
\newblock Citations: Co-citation in the scientific literature: A new measure of
  the relationship between two documents.
\newblock {\em Journal of the American Society for Information Science}, pages
  265--269, 1973.

\bibitem{Whittaker1989Creativity}
John Whittaker.
\newblock Creativity and conformity in science: Titles, keywords and co-word
  analysis.
\newblock {\em Social Studies of Science}, 19(3):473--496, August 1989.

\bibitem{Zitt2006Delineating}
Michel Zitt and Elise Bassecoulard.
\newblock Delineating complex scientific fields by an hybrid lexical-citation
  method: An application to nanosciences.
\newblock {\em Information Processing \& Management}, 42(6):1513--1531, 2006.

\end{thebibliography}

\bibliographystyle{plain}

\end{document}